\documentclass[aps, prc, twocolumn, tightenlines, letterpaper, amsmath, amssymb, preprintnumbers, superscriptaddress, floatfix, longbibliography, nofootinbib]{revtex4-2}

\usepackage[T1]{fontenc}
\usepackage{bm}
\usepackage{xspace}
\usepackage[dvipsnames]{xcolor}
\usepackage{graphicx}
\usepackage{tabularx}
\usepackage{enumitem}
\usepackage{braket}
\usepackage{dsfont}
\usepackage[normalem]{ulem}

\usepackage{amsmath, nccmath}
\usepackage{lipsum}
\usepackage{cancel}
\usepackage{bm}

\usepackage[hypertexnames=false]{hyperref}
\hypersetup{
    colorlinks=true,       
    linkcolor=blue,          
    citecolor=blue,        
    filecolor=blue,      
    urlcolor=blue           
}

\usepackage{orcidlink}

\newcommand{\Id}{\mathds{1}}

\newcommand{\MS}{\overline{\mathcal{M}}(\hat{\bf S})}
\newcommand{\ES}{\overline{\mathcal{E}}(\hat{\bf S})}

\def\si{{}^1\kern-.14em S_0}
\def\siii{{}^3\kern-.14em S_1}
\def\piii{{}^3\kern-.14em P_1}
\def\diii{{}^3\kern-.14em D_1}

\makeatletter 
    
\renewcommand\onecolumngrid{
\do@columngrid{one}{\@ne}%
\def\set@footnotewidth{\onecolumngrid}
\def\footnoterule{\kern-6pt\hrule width 1.5in\kern6pt}%
}

\renewcommand\twocolumngrid{
        \def\footnoterule{
        \dimen@\skip\footins\divide\dimen@\thr@@
        \kern-\dimen@\hrule width.5in\kern\dimen@}
        \do@columngrid{mlt}{\tw@}
}%

\makeatother  

\begin{document}

\title{Quantum Complexity Fluctuations from Nuclear and Hypernuclear Forces}

\author{Caroline E. P. Robin\,\orcidlink{0000-0001-5487-270X}}
\email{crobin@physik.uni-bielefeld.de}
\affiliation{Fakult\"at f\"ur Physik, Universit\"at Bielefeld, D-33615, Bielefeld, Germany}
\affiliation{GSI Helmholtzzentrum f\"ur Schwerionenforschung, Planckstra{\ss}e 1, 64291 Darmstadt, Germany}

\author{Martin J.~Savage\,\orcidlink{0000-0001-6502-7106}}
\email{mjs5@uw.edu}
\thanks{On leave from the Institute for Nuclear Theory.}
\affiliation{InQubator for Quantum Simulation (IQuS), Department of Physics, University of Washington, Seattle, WA 98195}

\date{\today}

\begin{abstract}
Toward an improved understanding 
of the role of
quantum information in nuclei 
and exotic matter,  
we examine the quantum magic (non-stabilizerness) in low-energy strong interaction processes.
As stabilizer states can be prepared efficiently using classical computers, 
and include classes of entangled states,
it is quantum magic and fluctuations in quantum magic, 
together with entanglement,
that determine 
computational resource requirements.
As a measure of  fluctuations in quantum magic,
and hence the severity of the exponentially-scaling classical computing resource requirements,
induced by scattering,
the  ``magic power'' of the S-matrix is introduced.
This provides indirect experimental constraints on quantum resources required to model nuclei and dense matter using fault-tolerant quantum computers. 
Using experimentally-determined
scattering phase shifts and mixing parameters,
the magic power in nucleon-nucleon and 
hyperon-nucleon scattering, 
along with the magic in the deuteron, 
are found to exhibit interesting and distinct features.
The $\Sigma^-$-baryon is identified as a potential candidate catalyst for enhanced spreading of magic and entanglement in dense matter, depending on in-medium decoherence.
\end{abstract}
\maketitle

\section{Introduction}

There has been tremendous progress in analyzing static and dynamical properties of 
quantum few-body and many-body systems from the point of view of quantum information.
One focus has been the characterization of entanglement features of such systems, to gain a better understanding of this phenomenon, its role in physical processes and connections to the 
fundamental laws of nature 
(for recent reviews, see Refs.~\cite{Banuls:2019bmf,Klco:2021lap,Bauer:2022hpo,Beck:2023xhh,Bauer:2023qgm,DiMeglio:2023nsa}).
In nuclear physics, this includes studies of entanglement at various energy scales, from systems relevant to QCD and high-energy phenomena (for example, Refs.~\cite{Beane:2019loz,Klco:2021cxq,Klco:2021biu,Gong:2021bcp,Mueller:2021gxd,Farrell:2022wyt,Mueller:2022xbg,Barata:2023jgd,Florio:2023mzk,Klco:2023ojt,Carena:2023vjc,Florio:2024aix,Farrell:2024mgu,Low:2024mrk}), to investigations of entanglement in few-baryon scattering~\cite{Beane:2018oxh,Beane:2020wjl,Beane:2021zvo,PhysRevC.107.025204,Bai:2022hfv,Bai:2023rkc,Bai:2023tey,Kirchner:2023dvg,Miller:2023snw,Miller:2023ujx}, to the structure of nuclei and nuclear models ~\cite{Johnson:2022mzk,PhysRevC.92.051303,Kruppa:2020rfa,Robin:2020aeh,Kruppa:2021yqs,Pazy:2022mmg,Tichai:2022bxr,Perez-Obiol:2023wdz,Gu:2023aoc,liu2023hints,Bulgac:2022cjg,Bulgac:2022ygo, PhysRevA.103.032426,Faba:2021kop,Faba:2022qop,Hengstenberg:2023ryt,Lacroix:2024drc,Bai:2023hrz,Bai:2024omg}, as well as dense neutrino systems (for example Refs.~\cite{Cervia:2019res,Patwardhan:2021rej,Lacroix:2022krq,Illa:2022zgu,Siwach:2022xhx,Roggero:2022hpy,Martin:2023ljq,Balantekin:2023qvm}). 

Importantly, without a classical analog, entanglement naturally appears as a key concept separating classical and quantum computations.
This has led to the development and re-interpretation of a range of methods that reorganize quantum many-body (QMB) problems around entanglement. 
One major example are tensor-network methods (for a recent review, see Ref.~\cite{Banuls2023}), such as the density-matrix renormalization group~\cite{PhysRevLett.69.2863}, with various adaptations in nuclear physics~\cite{PhysRevC.65.054319,Papenbrock_2005,PhysRevC.67.051303,PhysRevC.69.024312,PhysRevLett.97.110603,PhysRevC.79.014304,Dukelsky:2004vv,PhysRevC.78.041303,PhysRevC.88.044318,PhysRevC.106.034312,PhysRevC.92.051303,Tichai:2022bxr,Tichai:2024cyd}. In a related spirit, Refs.~\cite{GortonThesis,Johnson:2022mzk} introduced a reduced-basis method utilizing low proton-neutron entanglement. Approaches reorganizing entanglement via variational principles, with benefits for classical-quantum simulations of nuclear systems have also been developed~\cite{Robin:2020aeh,Robin:2023pgi}.

On the other hand, it is known from the Gottesman-Knill theorem~\cite{gottesman1998heisenberg} 
that stabilizer states, which include classes of entangled states, 
can be prepared efficiently using classical computers.
As such, entanglement measures alone are insufficient to assess 
the quantum resource requirements for simulating many-body systems,
and should be supplemented with measures of non-stabilizerness, or ``magic''~\cite{Aaronson_2004,Bravyi_2005,Stahlke_2014,Pashayan_2015,Bravyi_2016,Leone:2022lrn}. 
{While scaling polynomially with the number of gates from the classical gate set, 
the classical computing resources required to prepare a quantum state grow exponentially 
with the number of T-gates (or equivalent gates defining the universal quantum gate set), 
and hence with the magic in the state. Incidentally, T-gates are optimal for generating magic~\cite{LiLuo2022a}.

In this article, we initiate studies of magic in 
nuclei and dense matter by examining its role in low-energy nucleon-nucleon (NN) and hyperon-nucleon (YN) scattering.
This establishes a robust first step toward quantifying the fault-tolerant quantum computing resources that are required to prepare and evolve states of nuclei and hadronic matter, in terms of baryon degrees of freedom.
\footnote{These constraints are only abstractly related to the quantum resources required to perform QCD simulations of these systems.
As an illustration, if none of the S-matrix elements changed the magic between initial and final hadronic states, the ground state of the theory could be found efficiently using classical computation.
To be clear, the results presented in this work are not indicative of the resources required to compute scattering phase shifts from QCD, but do correspond to the impact 
of two-body scattering processes
on the computational resources required to simulate many-body systems.
}
It builds upon connections between entanglement suppression and emergent symmetries of the strong interactions~\cite{Beane:2018oxh,Low:2021ufv,PhysRevC.107.025204,Beane:2020wjl,liu2023hints},
and works in other arenas~\cite{Beane:2019loz,Beane:2021zvo,Carena:2023vjc}.
Using the work of Leone, Oliviero and Hamma in defining the magic power of a unitary operator~\cite{Leone_2022}, we investigate the magic power of the S-matrix in two-particle scattering channels that can be mapped to one and two qubits.
We find that magic patterns do not always follow entanglement patterns, in particular,
there are certain states that exhibit large entanglement and zero magic in specific energy regions, suggesting that the computational complexity of these processes could be energy dependent.
Interestingly, we find that the magic in the deuteron (induced by the tensor force) 
takes approximately the same value as the maximum magic power of the NN S-matrix.
While the magic power in $\Lambda$N scattering is found to remain small over a large range of energies, the magic power in $\Sigma^-n$ scattering rapidly reaches its maximum value which persists up to high energies.
This raises the intriguing possibility that $\Sigma^-$s may catalyze the growth of entanglement and magic in dense exotic matter.

\section{Definitions}

Formally, a $n$-qubit pure state $\ket{\Psi}$ is a stabilizer state if there exists a 
subgroup $\mathcal{S}(\ket{\Psi})$ of the Pauli group 
$\mathcal{G}_n  = \{ \varphi \, \hat{P}_1 \otimes \hat{P}_2  \otimes ... \otimes \hat{P}_n  \}$, 
where $\hat{P}_i \in \{ \Id, \sigma_x, \sigma_y, \sigma_z  \}$ and $\varphi \in \{\pm 1, \, \pm i \}$, with $|\mathcal{S}(\ket{\Psi})| =2^n$ elements, 
such that $\hat{P} \ket{\Psi} = \ket{\Psi}$ for all $\hat P \in \mathcal{S}(\ket{\Psi})$.  
The subgroup $\mathcal{S}(\ket{\Psi})$ is called the stabilizer group of $\ket{\Psi}$ and is Abelian~\cite{Gottesman_1996,gottesman1997stabilizer,Calderbank_1997,10.5555/2638682.2638691}. 
Stabilizer states can be prepared with stabilizer circuits, {\it i.e.} using Hadamard (H), phase (S) and CNOT gates (see appendix \ref{app:gates}). 
These Clifford gates alone are insufficient for universal quantum computation, 
which can be realized by including the non-Clifford T-gate.
Non-stabilizer states, or "magic states", 
which {\it a priori} cannot be efficiently prepared classically,
can be prepared using T-gates and Clifford gates.
Therefore, the resources required for quantum simulations
are given in terms of T-gate counts rather than CNOT-gate counts 
(while the later is currently relevant for the depth of quantum circuits 
that can be executed on Noisy Intermediate Scale Quantum (NISQ)-era~\cite{Preskill:2018jim} hardware).
While the formalism of stabilizer states has originally been developed 
for quantum error correction~\cite{Gottesman_1996,gottesman1997stabilizer,Calderbank_1997}, 
in the context of QMB physics, 
magic, together with entanglement, dictates the 
computational complexity.

Measures of magic based on Rényi entropies have been introduced~\cite{Leone_2022,Leone:2024lfr} building on some of the mathematical underpinnings from Refs.~\cite{kueng2015qubit,zhu2016clifford},
and a follow-up protocol to measure magic on a quantum processor was proposed and demonstrated in Ref.~\cite{Oliviero_2022}. 
Investigations of magic in matrix-product states have been performed in Refs.~\cite{Haug:2022vpg,Haug:2023hcs,frau2024nonstabilizerness,lami2024quantum}. Some recent works
developed computations of magic in the Ising model~\cite{Oliviero:2022euv,Rattacaso:2023kzm},
in two-dimensional lattice gauge theories~\cite{Tarabunga:2023ggd},
and in potential simulations of quantum gravity~\cite{Cepollaro:2024qln}. 
Also very recently, the use of doped stabilizer states has been proposed to represent energy eigenstates of certain QMB systems~\cite{Gu:2024qvn,gu2024doped}, 
and to develop efficient algorithms for their classical simulations.
Furthermore, it has been shown in a particular system 
that a phase transition in the scaling of magic occurs at a different measurement rate 
to that of entanglement~\cite{Fux:2023brx}.

Magic in a $n$-qubit state can be quantified by
considering a general expansion of the density matrix $\hat \rho$ in term of $n$-qubit Pauli strings
\begin{align}
    \hat \rho = \frac{1}{d} \sum_{\hat P \in \tilde{\mathcal{G}}_n } c_P \hat P \; , 
\end{align}
where $d=2^n$, $\tilde{\mathcal{G}}_n$ is the group of Pauli strings with phases $+1$, and $c_P =  Tr(\hat \rho \hat P)$. 
For a pure state $\hat \rho = \ket{\Psi}\bra{\Psi}$, where $c_P =\braket{\Psi |\hat P | \Psi}$, Ref.~\cite{Leone_2022} showed that $\Xi_P \equiv c_P^2/d$ is a probability distribution which can be interpreted as the probability for $\hat \rho$ to be in $\hat P$. 
\footnote{For mixed states, the $\Xi_P$ do not correspond to a probability distribution.  However, they can be used as re-scaling factors to define measures of magic~\cite{frau2024nonstabilizerness}. For an example, see appendix \ref{app:Sigma_cat}.}
The central step in quantifying magic comes from the demonstration 
that $\ket{\Psi}$ is a stabilizer state if and only if the coefficients $c_P = \pm 1$ for $d$ mutually commuting Pauli strings~\cite{zhu2016clifford} (and $c_P=0$ for the remaining $d(d-1)$ strings),
and thus, $\Xi_P = 1/d$ or $0$.
Consequently, Rényi entropies defined as
\begin{align}
    \mathcal{M}_\alpha(\ket{\Psi}) = -\log d + \frac{1}{1-\alpha} \log \left( \sum_P \Xi_P^\alpha \right)
\end{align}
provide a measure of magic, which vanishes for stabilizer states due to the added offset of 
$-\log d$~\cite{Leone_2022}.
While the present study can be carried out 
using any Rényi entropy convention, for consistency with previous works~\cite{Beane:2018oxh}, 
we focus on the linear entropy 
(1-Rényi entropy, or Shannon entropy, where the $\log$ is expanded to linear order)
\begin{align}
    \mathcal{M}(\ket{\Psi})  \equiv \mathcal{M}_{lin}(\ket{\Psi}) = 1 - d \sum_P \Xi_P^2 \; ,
    \label{eq:linear_magic_state}
\end{align}
which also vanishes for stabilizer states.

The starting point of our analysis of magic in scattering is to recognize that the 
action of the S-matrix on an initial stabilizer state can produce a final state that has magic.
A difference between entanglement and magic in this setting is that 
one qubit can be in a state with magic, but is obviously unentangled.
To describe magic in scattering processes, we introduce the magic power of the S-matrix, $\overline{\mathcal{M}}(\hat {\bf S})$, 
as the average magic induced by the operator $\hat {\bf S}$ on all $n$-qubit stabilizer states $\ket{\Psi_i}$:
\begin{align}
    \overline{\mathcal{M}}(\hat {\bf S}) \equiv \frac{1}{\mathcal{N}_{ss}} \sum_{i=1}^{\mathcal{N}_{ss}}  \mathcal{M} \left( \hat {\bf S} \ket{\Psi_i} \right) \; ,
\label{eq:Magic_Power}
\end{align}
where $\mathcal{N}_{ss}$ denotes the total number of $n$-qubit stabilizer states. 
This definition is analogous to the definition of the {\it entanglement power} of the S-matrix~\cite{Beane:2018oxh}
(a special case of the entangling power of a given unitary operator~\cite{PhysRevA.63.040304,BallardWu:2011}).
It is well known that a one-qubit system has $\mathcal{N}_{ss} = 6$ stabilizer states,
corresponding to the eigenstates of the Pauli operators $\sigma_x$, $\sigma_y$ and $\sigma_z$ 
(see appendix \ref{app:StabStates}),
\begin{align}
    &\ket{0}, \;  \ket{1}, \; \ket{+}=\frac{\ket{0}+\ket{1}}{\sqrt{2}}, \; \ket{-}=\frac{\ket{0}-\ket{1}} {\sqrt{2}}, \nonumber \\
    &\ket{+i}=\frac{\ket{0}+i\ket{1}}{\sqrt{2}}, \; \ket{-i}=\frac{\ket{0}-i\ket{1}}{\sqrt{2}} 
    \; .
    \label{eq:6stabs}
\end{align}
Two-qubit systems have 36 stabilizer states corresponding to tensor products of one-qubit stabilizer states, and 24 entangled stabilizers obtained by acting with CNOT gates
(see appendix \ref{app:StabStates})
which amount to a total of 60 stabilizer states.
This can be generalized to $n$ qubits using a recursive formula: $\mathcal{N}_{ss}(n)= 2(2^{n} + 1) \mathcal{N}_{ss}(n-1)$~\cite{10.5555/2638682.2638691}.

Similarly, the entanglement power of the S-matrix~\cite{Beane:2018oxh} 
can be redefined to
computing the average entanglement induced by $\hat {\bf S}$ over the tensor-product $n$-qubit stabilizer states $\ket{\Psi_i}$:
\begin{align}
    \overline{\mathcal{E}}(\hat {\bf S}) \equiv \frac{1}{\mathcal{N}_{ss}^{TP}} \sum_{i=1}^{\mathcal{N}_{ss}^{TP}}  \mathcal{E} \left( \rho_i^{(1)}(\hat {\bf S}) \right) \; ,
\label{eq:Entang_Power}
\end{align}
where $\mathcal{N}_{ss}^{TP}$ is the number of tensor-product stabilizer states, and 
$\rho_i^{(1)}(\hat {\bf S}) = \mbox{Tr}_2 \left[ \rho_i^{(12)}(\hat {\bf S}) \right]$  is the outgoing reduced density matrix for particle 1, obtained by tracing the full outgoing density matrix $\rho_i^{(12)}(\hat {\bf S}) = \hat {\bf S} \ket{\Psi_i}\bra{\Psi_i} \hat {\bf S}^\dagger$ over particle 2.
This definition recovers the results obtained by continuous integration over spin orientations of initial tensor-product states~\cite{Beane:2018oxh}.
\footnote{This (surprising) equality between the result obtained by continuous integration and by summing over the tensor-product stabilizer states is due to the fact that averaging the fluctuations in entanglement independently over the two Bloch spheres is the same is averaging it over the basis vectors of 
$\hat X$, $\hat Y$ and $\hat Z$ of each qubit.  
These basis vectors are obtained by applications of the Hadamard gate and the product of a Hadamard and Phase gate, which transform
$|0\rangle, |1\rangle$ into $|+\rangle, |-\rangle$ and $|+i\rangle, |-i\rangle$ given in Eq.~(\ref{eq:6stabs}), respectively.
}

\section{The Magic of the Deuteron as a Single Qubit}

We first consider  
neutron-proton ($np$) 
scattering in the $J=1$ $\siii$-$\diii$ 
coupled channels, which contains the deuteron bound state. 
Suppressing individual spin indices,
this is a two-component system which can be mapped to one qubit with basis states 
$\ket{\siii} \equiv \ket{0}$ and $\ket{\diii} \equiv \ket{1}$.
With this mapping, the S-matrix, using the Stapp parametrization~\cite{PhysRev.105.302}, 
is:
\begin{eqnarray}
     S_{(J=1)} & = & 
    \left(
    \begin{array}{cc}
        e^{ i \overline{\delta}_{0}} & 0\\
        0 & e^{ i \overline{\delta}_{2}}
    \end{array}
    \right)
    \left(
    \begin{array}{cc}
        \cos 2\overline{\epsilon}_1 & i \sin 2\overline{\epsilon}_1 \\
        i \sin 2\overline{\epsilon}_1 & \cos 2\overline{\epsilon}_1
    \end{array}
    \right) \nonumber \\
    &&\times
    \left(
    \begin{array}{cc}
        e^{ i \overline{\delta}_{0}} & 0\\
        0 & e^{ i \overline{\delta}_{2}}
    \end{array}
    \right)
    \ \ ,
    \label{eq:Smatrix_Stapp}
\end{eqnarray}
where $\overline\delta_{0}$ and $\overline\delta_{2}$ are the phase shifts for the $\siii$ and $\diii$ waves, respectively, and $\overline{\epsilon}_1$ is the mixing angle (for another parametrization, see appendix \ref{app:1qubitSmatBB}).
Acting on each of the six stabilizer states associated with one qubit given in Eq.~(\ref{eq:6stabs}),
the magic power of the S-matrix determined using Eq.~\eqref{eq:Magic_Power} is
\begin{eqnarray}
\overline{\cal M}(\hat {\bf S}_{(J=1)}) & = &  
{1\over 6}
\Bigl[\ 
\sin^2(8 \overline{\epsilon}_1) 
\ +\ 
\cos^8(2 \overline{\epsilon}_1) \sin^2 (4 \Delta\overline\delta)
\nonumber \\
&& + \ {7\over 4} \sin^4(4 \overline{\epsilon}_1) \sin^2 (2 \Delta\overline\delta)
\Bigr]
\label{eq:sdCoupledMagicStapp}
\ \ ,
\end{eqnarray}
where $\Delta{\overline\delta} \equiv \overline\delta_0 - \overline\delta_2$.
\begin{figure}[!ht]
    \centering
    \includegraphics[width=\columnwidth]{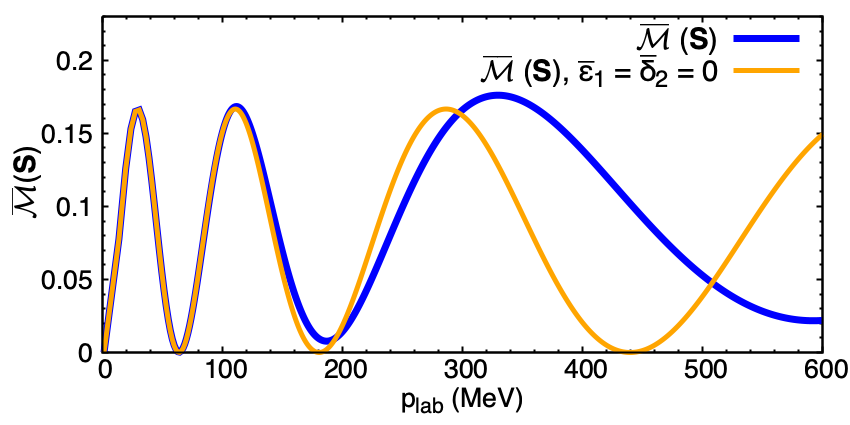}
    \caption{
    The magic power $\overline{\mathcal{M}}({\hat {\bf S}}_{(J=1)})$ in $np$ scattering in the 
    $\siii$-$\diii$  coupled channels determined using Eq.~\eqref{eq:Magic_Power}, as a function of laboratory momentum p$_{\rm lab}$. 
    The blue curve shows the full result using the Nijm93 phase-shift analysis~\cite{PhysRevC.49.2950,NNonline}, 
    while the orange curve corresponds to the limit $\overline{\epsilon}_1 =\overline{\delta}_2=0 $. 
    }
    \label{fig:1qb_deuteron_magic}
\end{figure}
Figure~\ref{fig:1qb_deuteron_magic} shows 
$\overline{\cal M}(\hat {\bf S}_{(J=1)})$ using  
$\overline{\delta}_0, \overline{\delta}_2, 
\overline{\epsilon}_1$
from the Nijm93 fit to $np$ scattering data~\cite{NNonline}
as a function of momentum in the laboratory frame~\footnote{
The quality of the experimental NN scattering data highly 
constrains phenomenological phase-shift analyses, 
thus we restrict ourselves to one such NN potential, 
Nijm93~\cite{PhysRevC.49.2950}, for demonstrative purposes. }.
The magic power exhibits significant structure over a small energy range near threshold, due to the rapidly varying phase shifts near unitarity.
The minima of the magic power are found near 
p$_{\rm lab} \approx 0, 64, 187, ...$ MeV, 
and 
maxima of $\approx 0.17$ 
near p$_{\rm lab}\approx 31, 110, 330 ...$ MeV.
These maximum values are to be compared with the maximum possible value for one-qubit magic, which is $1/3$.

It is interesting to calculate the magic in the deuteron, the loosely-bound $J=1$ $np$ 
ground state in the 
$\siii$-$\diii$ coupled channels,
using the mapping described above
$\ket{\psi}_{\text{deuteron}} = A_S \ket{\siii} + A_D \ket{\diii}$. 
As
the aysmptotic D/S-ratio is
a useful parameterization for describing scattering data, the
Nijm93 potential provides a D-wave amplitude of $A_D \approx 0.24$, corresponding to a probability of $\approx 5.8\%$~\cite{PhysRevC.49.2950}. Using this value, the linear magic in Eq.~\eqref{eq:linear_magic_state} takes the value 
$\mathcal{M}(\ket{\psi}_{\text{deuteron}})\approx0.17$,
which is intriguingly 
close to the maxima of magic power in this channel
shown in Fig.~\ref{fig:1qb_deuteron_magic}. 
This suggests that there may be a connection between the magic power 
in the continuum and 
the magic in bound states,  but this is merely speculation.
For the deuteron, 
the magic is generated by the tensor force that provides the mixing between $\siii$ and
$\diii$ channels. 
The maximum value $\mathcal{M}^{max}=0.25$ (for a real wave function) 
would have been obtained for a nearby D-state 
amplitude of $A_D=\sin {\pi\over 8}$ and hence a 
probability of $\approx 14.6 \%$.

\section{The Magic in NN and YN scattering}

Let us now turn to the magic and entanglement in the spin-sector of S-wave NN scattering 
(and neglect the mixing with the D-wave).
The nucleons can be reduced to their spin degrees of freedom, 
and mapped onto two qubits, 
with basis states 
$\ket{0}_{\rm N} = \ket{\uparrow}_{\rm N}$ and $\ket{1}_{\rm N} = \ket{\downarrow}_{\rm N}$ (N$=n,p$). 
The S-matrix in these channels is,
\begin{align}
    \hat {\bf S} = \frac{1}{4} \left( 3 \ e^{2 i \delta_1} +  e^{2 i \delta_0} \right ) \hat\Id 
    + \frac{1}{4} \left( \ e^{2 i \delta_1} -  e^{2 i \delta_0} \right ) \hat{\boldsymbol{\sigma}} . \hat{\boldsymbol{\sigma}} \; ,
\label{eq:Smatrix_2q}    
\end{align}
which is a combination of the identity and spin-exchange operator (SWAP gate)~\cite{Beane:2018oxh}. 
In Eq.~\eqref{eq:Smatrix_2q}, 
$\delta_0$ and $\delta_1$ are the phase shifts associated with the 
$\si$ and $\siii$ channels, respectively.
Using Eq.~\eqref{eq:Magic_Power} and Eq.~\eqref{eq:Entang_Power} to compute the magic power   and entanglement power   of the S-matrix, we obtain
\begin{eqnarray}
\MS &=& {3\over 20} \Bigl( 3 + \cos(4\, \Delta \delta) \Bigr) \sin^2(2\, \Delta \delta) \; , \\
\ES & = & {1\over 6} \sin^2 (2 \, \Delta \delta ) \; ,
\label{eq:Magic_Entang_Power_NN_2q}
\end{eqnarray}
where $\Delta \delta \equiv \delta_1-\delta_0$.
Interestingly, the magic power differs 
in form
from the entanglement power by the extra cosine term.
Both $\ES$ and $\MS$ vanish for $\Delta \delta= k \pi/2$ ($k$ integer), 
which is encountered in the case of SU(4) symmetry ($\delta_1 = \delta_0$)
that emerges in the large-N$_c$ limit of QCD~\cite{Kaplan:1995yg}, 
and for some special fixed points forming the Klein group~\cite{Beane:2018oxh}. 
Both present maxima $\MS_{max} = 0.3$ and $\ES_{max} = 1/6 \approx 0.167$ at 
$\Delta \delta = (k+1/2) \pi /2$. The magic power, however, presents a small plateau around this value.
\begin{figure}[!ht]
    \centering
    \includegraphics[width=\columnwidth]{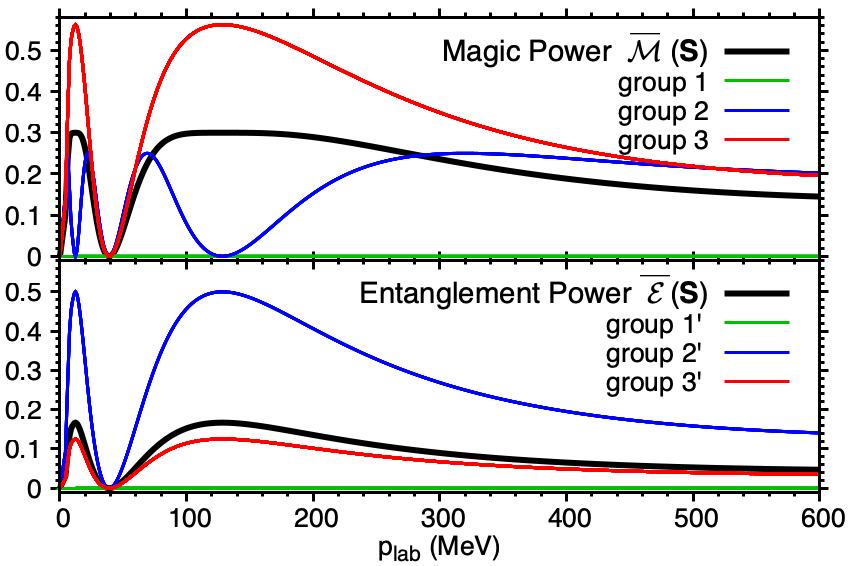}
    \caption{The magic power $\MS$ (left panel) and entanglement power $\ES$ (right panel) of the S-matrix in the spin degrees of freedom in $np$ scattering as a function of the momentum p$_{\rm lab}$ in the laboratory frame. The phase shifts are those from the Nijm93 parametrization \cite{PhysRevC.49.2950}, obtained from Ref.~\cite{NNonline}. 
    The groups $i'$ are restricted to the tensor-product states of groups $i$, respectively.}
    \label{fig:NN_magic_entang}
\end{figure}
The magic and entanglement power of the S-matrix computed with Nijm93 phase shifts are shown in Fig.~\ref{fig:NN_magic_entang} (black curves).
Interestingly, it is seen that 
$\MS$ overall is larger than $\ES$ by a factor $\approx 2$ over the full energy range. 
$\MS$ roughly follows the trend of $\ES$, except in the region around 
p$_{\rm lab}\approx 100-200$ where $\MS$ presents a plateau.
To underpin the origin of this plateau we have examined the individual contribution $\mathcal{M} (\hat{\bf S} \ket{\Psi_i})$ of each initial stabilizer state $\ket{\Psi_i}$. 
We find that each of them can be classified in one of three groups of states which contribute in the same way to the magic or entanglement power.
These groups are detailed in  
the appendix \ref{app:zooming}, 
and  each contain tensor-product and entangled states. 
Their contributions are shown with green, blue and red curves in Fig.~\ref{fig:NN_magic_entang}.
It is seen that the stabilizers that generate the largest amount of entanglement do not coincide with those generating the largest magic. 
In particular, "group 2" presents a distinct behaviour around p$_{\rm lab} \approx 128$ MeV (which corresponds to $p^* \approx 64$ MeV in the center-of-mass frame), 
where the magic cancels but the entanglement takes its maximal value. 
The physical meaning of this dip in the magic remains unclear, 
but coincidentally, this region corresponds to the start of the t-channel cut at $p^* = m_\pi/2$, 
beyond which the effective range expansion is no longer valid.
Conversely, there are no energies at which the system is unentangled and magic.
\\
\\

The above analysis is also applied 
to YN scattering. 
Specifically, we consider $\Sigma^- n$ and $\Lambda p$ 
which may have  importance for the structure of dense matter, 
as formed in core-collapse supernova (for a recent discussions, see Ref.~\cite{Baym:2017whm,Blacker:2023opp,Mu:2023yqh}).
For these processes, 
the phase shifts derived from chiral effective field theory 
($\chi$EFT)\cite{Savage:1995kv}
at next-to-next-to leading order (N2LO) are adopted~\cite{Haidenbauer:2023qhf}
(for a comparison to results with phenomenological phase shifts, see appendix \ref{app:comp_EFT_NSC97}).
Figure~\ref{fig:YN_magic_entang} shows the resulting 
magic and entanglement power of the $\Lambda p$ 
and 
$\Sigma^- n$  S-matrices.
\begin{figure}[!ht]
    \centering
    \includegraphics[width=\columnwidth]{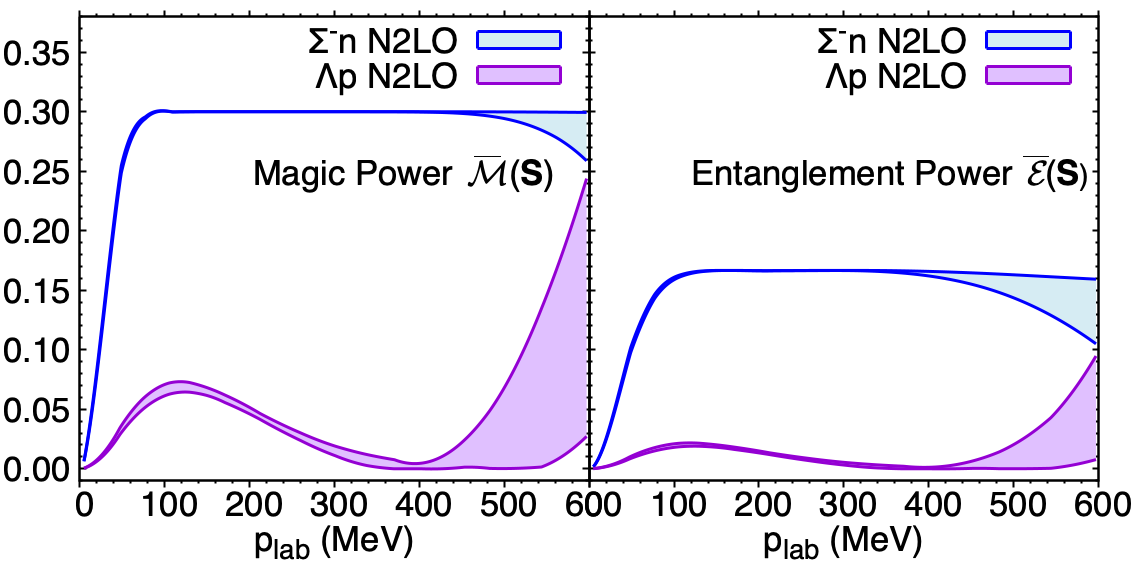}
    \caption{Magic power $\MS$ (left panel) and entanglement power $\ES$ (right panel) in $\Sigma^-n$ and $\Lambda$p scattering, obtained using N2LO-$\chi$EFT phase shifts from Ref.~\cite{Haidenbauer:2023qhf}.
    We have assumed isospin symmetry between $\Sigma^+p$ and $\Sigma^-n$, and neglected Coulomb interactions.
    The uncertainty bands represent the maximum and minimum values in magic and entanglement derived from the N2LO phase-shift uncertainty bands~\cite{Haidenbauer:2023qhf}.}
    \label{fig:YN_magic_entang}
\end{figure}
We observe significant differences compared to the $np$ channel, 
due to dissimilarities in the behavior of the phase shifts.
In particular, 
$\Delta \delta$ varies only slowly
in $\Sigma^- n$ scattering,  and takes values close to $\pi/4$ 
over an extended range of energies, which makes $\MS$ and $\ES$ almost constantly maximal.
Conversely, in $\Lambda p$ scattering, the phase shifts $\delta_0$ and $\delta_1$ take comparable values, which largely suppresses magic and entanglement powers.  
This may be interpreted as due to the nature of the spin of the hyperon.  
In the case of the $\Lambda$, the spin is carried mostly by the s-quark, and the magic and entanglement power suggest that it is largely decoupled from the other spin dynamics.
In contrast, the spin of the $\Sigma^-$
is carried by both the s-quark and the d-quarks, which appear to be strongly coupled to the neutron, and able to provide substantial fluctuations in entanglement and magic over a wide range of energies.
Using a 3-qubit example and assuming pure-state evolution,
we compute that two successive 
$\Sigma^-n$ scatterings induce a linear magic of
$\langle M \rangle \approx 0.405$ in the two neutrons  over
a significant kinematic range in the absence of decoherence
(see appendix \ref{app:Sigma_cat} for details). 
While grossly simplified model of dense matter,
we speculate that this could provide a mechanism to grow and spread magic and entanglement in dense matter,
via $\Sigma^-$-catalysis.
Of course, critical to this discussion is the assumed quantum coherence between scatters, which is sure to be degraded in a non-equilibrium astrophysical setting. 
The degree of coherence remains to be determined.

Preparing and evolving one-qubit and two-qubit wavefunctions is easy to do with a classical computer. 
However one may wonder if, for example, the one-qubit magic computed in the $\siii$-$\diii$ coupled channels
reflects the long-distance magic, 
and thus the complexity of simulating $np$ scattering and the deuteron 
at the level of quarks and gluons using lattice QCD techniques (see, for example, Refs.~\cite{Beane:2006mx,NPLQCD:2011naw,NPLQCD:2012mex,Yamazaki:2012hi,Orginos:2015aya,Yamazaki:2015asa,Wagman:2017tmp,Drischler:2019xuo,Davoudi:2020ngi,Horz:2020zvv,Amarasinghe:2021lqa,Aoki:2023qih,Detmold:2024iwz}).   
It would be remarkable if it did,
and
one of many challenges that remain is to understand how magic and entanglement evolve through the confinement scale~\cite{Florio:2023mzk}.
Our study indicates that the computational complexity for simulating NN scattering depends upon the energy. Performing such simulations at parameters corresponding to 
vanishing magic would then be computationally less demanding 
(and efficient for classical computers) than at points with local maxima in magic. 
To further address this question, more systematic studies of the scaling of magic with system sizes, and of how magic is affected by qubit mappings to fundamental or emergent degrees of freedom, are needed.

\section{Summary}

To summarize, as a step towards investigating 
the role of quantum magic in nuclear phenomena, and the associated 
computational complexity of their simulation, 
we have considered the 
magic power of the S-matrix to explore 
fluctuations in magic induced by 
low-energy
NN and YN scattering.
It is the magic in a large-scale entangled quantum state that 
signifies exponentially-scaling classical resource requirements to prepare the state
and hence the need for quantum computation at scale.
Using available phase-shift analyses, including N2LO YN interactions 
in a chiral expansion,
the tensor force is found to be responsible for interesting behavior of
the magic in the $\siii$-$\diii$ NN coupled channels, 
and the spin-spin interaction induces similarly interesting behavior 
in the spin degrees of freedom in the $\si$ and $\siii$ channels.
There is striking behavior in the magic power of the $\Sigma^- n$ S-matrix,
being approximately maximal and independent of energy over a large interval,
that could be relevant for evolution of exotic matter.
If the density of $\Sigma^-$s becomes appreciable, the results presented in this work indicate that their scattering processes may provide a significant contribution to the 
computational complexity of simulating such systems.

Systems that have  magic without entanglement, or entanglement without magic,
can be prepared efficiently using classical computers.
Therefore, both  magic and entanglement are important to consider 
in estimating the resources required for quantum simulations of  many-body systems 
of nucleons and hyperons.

\begin{acknowledgements}
\noindent
We would like to thank Emanuele Tirrito for his inspiring presentation at the IQuS workshop {\it Pulses, Qudits and Quantum Simulations}\footnote{\url{https://iqus.uw.edu/events/pulsesquditssimulations/}}, 
co-organized by Yujin Cho, Ravi Naik, Alessandro Roggero and Kyle Wendt, and for subsequent discussions,
an also related discussions with Alessandro Roggero and Kyle Wendt.
We also thank Ulf-G. Mei{\ss}ner and Johann Haidenbauer for discussions regarding the 
EFT analyses of YN interactions, 
and for sharing YN phase shifts derived from chiral EFT, 
as well as Sanjay Reddy for discussions related to the role of hyperons in dense matter.
This work was supported, in part,  by Universit\"at Bielefeld and ERC-885281-KILONOVA Advanced Grant (Caroline), by U.S. Department of Energy, Office of Science, Office of Nuclear Physics, InQubator for Quantum Simulation (IQuS)\footnote{\url{https://iqus.uw.edu}} under Award Number DOE (NP) Award DE-SC0020970 via the program on Quantum Horizons: QIS Research and Innovation for Nuclear Science\footnote{\url{https://science.osti.gov/np/Research/Quantum-Information-Science}} (Martin).
This work was supported, in part, through the Department of Physics\footnote{\url{https://phys.washington.edu}}
and the College of Arts and Sciences\footnote{\url{https://www.artsci.washington.edu}} at the University of Washington. 
We have made extensive use of Wolfram {\tt Mathematica}~\cite{Mathematica}.
\end{acknowledgements}

\bibliography{bibi_magic}

\onecolumngrid
\appendix
\section{Gates}
\label{app:gates}
\noindent
Here we present the gates used in the discussions in the main text.
Quantum circuits that can be efficiently simulated using classical computers are those involving only 
Clifford gates, The single-qubit H-gate and S-gate, and the two-qubit CNOT$_{ij}$-gate 
(a two-qubit control-X entangling gate where $i$ denotes the control qubit and $j$ the target qubit), 
given by, for example,
\begin{eqnarray}
{\rm H} & = & 
{1\over\sqrt{2}}\ \left(
\begin{array}{cc}
1&1\\ 1&-1
\end{array}
\right)
\ \ ,\ \ 
{\rm S}\ =\ 
\left(
\begin{array}{cc}
1&0\\ 0&i
\end{array}
\right)
\ \ ,\ \ 
{\rm CNOT}_{12} \ =\ 
\left(
\begin{array}{cccc}
1&0&0&0\\ 
0&1&0&0\\ 
0&0&0&1\\ 
0&0&1&0 
\end{array}
\right)
\ \ \ .
\end{eqnarray}
Repeated applications of this gate set 
\{ H, S, CNOT$_{ij}$ \}
to a $n$-qubit tensor-product state will generate the complete set of stabilizer states.
Inclusion of the T-gate,
\begin{eqnarray}
    {\rm T} & = &  
\left(
\begin{array}{cc}
1&0\\ 0&e^{i \pi/4}
\end{array}
\right)
\ \ ,
\end{eqnarray}
yields a complete gate set for universal quantum computation, 
which that repeated application of 
\{~H,~T,~CNOT$_{ij}$~\}
provides access to any circuit that can be simulated using a quantum computer.
As T-gates are a costly resource, 
one typically thinks about the gate set
\{ H, S, CNOT$_{ij}$, T \}, but keeping in mind that T$^2$=S.

\section{Stabilizer States}
\label{app:StabStates}
\noindent
As stated in the main text, 
a $n$-qubit pure state $\ket{\Psi}$ is said to be a stabilizer state if there exists a subgroup 
$\mathcal{S}(\ket{\Psi})$ of the Pauli group $\mathcal{G}_n  = \{ \varphi \, \hat{P}_1 \otimes \hat{P}_2  \otimes ... \otimes \hat{P}_n  \}$, where $\hat{P}_i \in \{ \Id, \sigma_x, \sigma_y, \sigma_z  \}$ and $\varphi \in \{\pm 1, \, \pm i \}$, with $|\mathcal{S}(\ket{\Psi})| =2^n$ elements, such that $\hat{P} \ket{\Psi} = \ket{\Psi}$ for all $\hat P \in \mathcal{S}(\ket{\Psi}) \subset \mathcal{G}_n$.  
The stabilizers associated with a given system are conveniently determined by 
a Pauli decomposition of the density matrix,
\begin{eqnarray}
    \rho & = & |\psi\rangle\langle\psi|\ =\ 
    {1\over d} \sum_{P \in \tilde{\mathcal{G}}_n} c_P \hat P
    \ ,\
    c_P \ =\  {\rm Tr}\left[ \rho . \hat P\right]
    \ ,
\end{eqnarray}
where $\tilde{\mathcal{G}}_n$ is the group of Pauli strings with phases $\varphi=+1$ only, 
$d=2^n$ and ${1\over d}\sum_P c_P^2 = 1$.
The stabilizers are associated with the set of coefficients containing $d$ coefficients $c_P$ with values $\pm 1$, and the others with values $c_P=0$.
A brute force way to generate the stabilizer states for an $n$-qubit state is to start in the 
$|0\rangle^{\otimes n}$ tensor-product state,  
exhaustively apply the Clifford gates, 
\{ H, S, CNOT$_{ij}$ \},  in all possible ways, and retain the distinct states that result.
This method works well for small systems, but the number of states grows exponentially with 
increasing $n$~\cite{10.5555/2638682.2638691} and soon becomes unmanageable. In that case, statistical sampling over such circuits 
provides a path forward.

For one qubit (with $d= 2$), there are 6 stabilizer states that have 2 Pauli operators satisfying 
$\hat{P} \ket{\Psi} = \ket{\Psi}$, which are listed in Table~\ref{tab:OneQstabs}.
\begin{table}[!htb]
\centering
\begin{tabularx}{0.23\columnwidth}{c|c} 
\hline\hline
$P$ & $|\psi\rangle$\\
\hline
$\hat \Id$, $\hat Z$ & $|0\rangle$ \\
$\hat \Id$, -$\hat Z$ & $|1\rangle$ \\
$\hat \Id$, $\hat X$ & $\ket{+} \equiv$ ${1\over\sqrt{2}} \left( |0\rangle+|1\rangle \right)$ \\
$\hat \Id$, -$\hat X$ & $\ket{-} \equiv$ ${1\over\sqrt{2}} \left( |0\rangle-|1\rangle \right)$ \\
$\hat \Id$, $\hat Y$ & $\ket{+i} \equiv$ ${1\over\sqrt{2}} \left( |0\rangle+i |1\rangle \right)$ \\
$\hat \Id$, -$\hat Y$ & $\ket{-i} \equiv$ ${1\over\sqrt{2}} \left( |0\rangle-i |1\rangle \right)$ \\
\hline\hline
\end{tabularx}
\caption{
One-qubit stabilizer states and their stabilizers. 
}
\label{tab:OneQstabs}
\end{table}
For two qubits (with $d=4$), there are 4 stabilizers for each of the 
60 stabilizer states given in Table~\ref{tab:TwoQstabs}.
36 of these states are tensor products formed from two of the one-qubit stabilizer states, while
the remaining 24 are entangled states of the two qubits.
 \begin{table}[!htb]
 \centering
 \begin{tabularx}{0.5\columnwidth}{c|cccc||c|cccc} 
 \hline\hline
 state & $|00\rangle$ & $|01\rangle$ & $|10\rangle$ & $|11\rangle$
 & state & $|00\rangle$ & $|01\rangle$ & $|10\rangle$ & $|11\rangle$\\
 \hline
 1 & 1 & 1 & 1 & 1    & 37 & 0 & 1 & 1 & 0 \\
 2 & 1 & -1 & 1 & -1  & 38 & 1 & 0 & 0 & -1\\
 3 & 1 & 1 & -1 & -1  & 39 & 1 & 0 & 0 & 1 \\
 4 & 1 & -1 & -1 & 1  & 40 & 0 & 1 & -1 & 0\\
 5 & 1 & 1 & i & i    & 41 & 1 & 0 & 0 & i\\
 6 & 1 & -1 & i & -i  & 42 & 0 & 1 & i & 0 \\
 7 & 1 & 1 & -i & -i  & 43 & 0 & 1 & -i & 0 \\
 8 & 1 & -1 & -i & i  & 44 & 1 & 0 & 0 & -i\\
 9 & 1 & 1 & 0 & 0    & 45 & 1 & 1 & 1 & -1\\
 10 & 1 & -1 & 0 & 0  & 46 & 1 & 1 & -1 & 1\\
 11 & 0 & 0 & 1 & 1   & 47 & 1 & -1 & 1 & 1\\
 12 & 0 & 0 & 1 & -1  & 48 & 1 & -1 & -1 & -1\\
 13 & 1 & i & 1 & i   & 49 & 1 & i & 1 & -i\\
 14 & 1 & -i & 1 & -i & 50 & 1 & i & -1 & i\\
 15 & 1 & i & -1 & -i & 51 & 1 & -i & 1 & i\\
 16 & 1 & -i & -1 & i & 52 & 1 & -i & -1 & -i \\
 17 & 1 & i & i & -1  & 53 & 1 & 1 & i & -i \\
 18 & 1 & -i & i & 1  & 54 & 1 & 1 & -i & i \\
 19 & 1 & i & -i & 1  & 55 & 1 & -1 & i & i \\
 20 & 1 & -i & -i & -1& 56 & 1 & -1 & -i & -i   \\
 21 & 1 & i & 0 & 0   & 57 & 1 & i & i & 1 \\
 22 & 1 & -i & 0 & 0  & 58 & 1 & i & -i & -1 \\
 23 & 0 & 0 & 1 & i   & 59 & 1 & -i & i & -1 \\
 24 & 0 & 0 & 1 & -i  & 60 & 1 & -i & -i & 1 \\
 25 & 1 & 0 & 1 & 0  \\
 26 & 0 & 1 & 0 & 1  \\
 27 & 1 & 0 & -1 & 0  \\
 28 & 0 & 1 & 0 & -1  \\
 29 & 1 & 0 & i & 0  \\
 30 & 0 & 1 & 0 & i  \\
 31 & 1 & 0 & -i & 0  \\
 32 & 0 & 1 & 0 & -i  \\
 33 & 1 & 0 & 0 & 0  \\
 34 & 0 & 1 & 0 & 0  \\
 35 & 0 & 0 & 1 & 0  \\
 36 & 0 & 0 & 0 & 1  \\
 \hline\hline
 \end{tabularx}
 \caption{
 The complete set of 60 two-qubit stabilizer states.  
 The left set are from the tensor product of one-qubit stabilizer states, 
 while the right set are entangled states.   
 These states are (generally) unnormalized, and require coefficients of either 1 or ${1\over\sqrt{2}}$ or ${1\over 2}$.
 }
 \label{tab:TwoQstabs}
 \end{table}
Finally, for three qubits, there are $1080$ stabilizer states 
(which are given in Ref.~\cite{10.5555/2638682.2638691}), 
and for four qubits there are $36 720$ stabilizer states~\cite{10.5555/2638682.2638691}.

\section{Nucleon-Nucleon Coupled Channels in the Blatt-Biedenharn Parameterization~\cite{PhysRev.86.399}}
\label{app:1qubitSmatBB}
\noindent
The S-matrix for scattering in the $\siii-\diii$ $J=1$ coupled channels
can be written in terms of two phase shifts and one mixing angle.
In the main text, the Stapp convention~\cite{PhysRev.105.302} was used to define the S-matrix, 
but others can be used, for example, the  Blatt-Biedenharn (BB) convention~\cite{PhysRev.86.399} 
(for a discussion, see Ref.~\cite{deSwart:1995ui}), defined by
\begin{eqnarray}
    \hat S_{(J=1)} & = & 
    \left(
    \begin{array}{cc}
        \cos\epsilon_1 & -\sin\epsilon_1 \\
        \sin\epsilon_1 & \cos\epsilon_1
    \end{array}
    \right)
    \left(
    \begin{array}{cc}
        e^{2 i \delta_{1\alpha}} & 0\\
        0 & e^{2 i \delta_{1\beta}}
    \end{array}
    \right)
    \left(
    \begin{array}{cc}
        \cos\epsilon_1 & \sin\epsilon_1 \\
        -\sin\epsilon_1 & \cos\epsilon_1
    \end{array}
    \right)
    \ \ .
\end{eqnarray}
The same procedure is used to determine the magic power of the S-matrix in this parameterization,
using the stabilizer states 
\begin{align}
    &\ket{0}, \;  \ket{1}, \; \ket{+}=\frac{\ket{0}+\ket{1}}{\sqrt{2}}, \; \ket{-}=\frac{\ket{0}-\ket{1}} {\sqrt{2}}, \nonumber \\
    &\ket{+i}=\frac{\ket{0}+i\ket{1}}{\sqrt{2}}, \; \ket{-i}=\frac{\ket{0}-i\ket{1}}{\sqrt{2}} 
    \; .
    \label{eq:6stabs_supp}
\end{align}
The magic power defined as
\begin{align}
    \overline{\mathcal{M}}(\hat {\bf S}) \equiv \frac{1}{\mathcal{N}_{ss}} \sum_{i=1}^{\mathcal{N}_{ss}}  \mathcal{M} \left( \hat {\bf S} \ket{\Psi_i} \right) \; ,
\label{eq:Magic_Power_suppl}
\end{align}
is found to be, in the BB convention,
\begin{eqnarray}
\overline{\cal M}(\hat S_{(J=1)}) & = &  
{1\over 384}
\left[
\left(
322 
+ 397 \cos (2 \Delta\delta)
+ 206 \cos (4 \Delta\delta)
+ 99 \cos (6 \Delta\delta)
\right) \sin^2 \Delta\delta
\right. \nonumber\\
& & \left.
\ -\ 
32\cos(16\epsilon_1) \sin^8 \Delta\delta
\ -\ 
56\cos(8\epsilon_1) \sin^4 (2 \Delta\delta)
\right]
\label{eq:sdCoupledMagic}
\ \ ,
\end{eqnarray}
where
$\Delta\delta = \delta_\alpha-\delta_\beta$.
We see that 
$\overline{\cal M}(\hat S_{(J=1)})=0$ when $\Delta\delta=0$, as expected, as the S-matrix becomes the identity, and interestingly, the contribution from $\epsilon_1$ is suppressed by higher orders in
$\Delta\delta$.
Phenomenologically  $\delta_\beta , \epsilon_1 \approx 0$ at low energies, 
and in this limit the magic power and entangling power are,
\begin{eqnarray}
\overline{\cal M}(\hat S_{(J=1)}) & = &  
{1\over 6} \sin^2 (4 \delta_\alpha )
\ \ ,\ \ 
{\cal E}(\hat S) \ =\  {1\over 4} \sin^2 (4 \delta_\alpha)
\ \ .
\end{eqnarray}
%

\section{Contributions from Stabilizer States}
\label{app:zooming}

\noindent
It is informative to examine the contributions $\mathcal{M} (\hat{\bf S} \ket{\Psi_i})$ of individual stabilizer states $\ket{\Psi_i}$ to the magic power of the S-matrix.
In the case of $np$ scattering in the coupled $\siii-\diii$ channels (deuteron), which was mapped to one qubit, there are contributions from the six stabilizer states in Table~\ref{tab:OneQstabs}. Their contributions to the magic power 
are shown in Fig.~\ref{fig:1qb_deuteron_magic_contrib}.
\begin{figure}[!ht]
    \centering
    \includegraphics[width=.45\columnwidth]{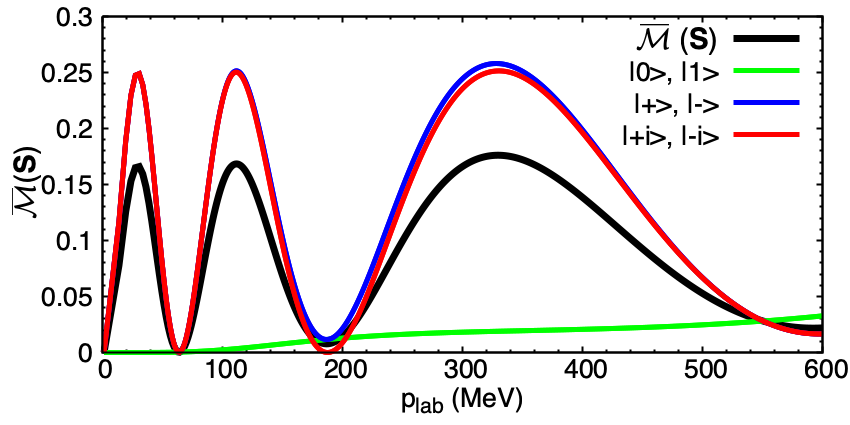}
    \caption{ Contributions of the six one-qubit stabilizer states to the magic power $\overline{\mathcal{M}}({\hat {\bf S}}_{(J=1)})$ in $np$ scattering in the 
    $J=1$ coupled $\siii$-$\diii$ channels as a function of laboratory momentum p$_{\rm lab}$, obtained with the Nijm93 phase shifts~\cite{PhysRevC.49.2950,NNonline}.
    }
    \label{fig:1qb_deuteron_magic_contrib}
\end{figure}
At low energies, the channels mix minimally,
and the S-matrix is approximately diagonal. In that limit, the stabilizer states $\ket{\siii}\equiv\ket{0}$ and $\ket{\diii}\equiv\ket{1}$ 
do not contribute magic, 
as they only acquire a global phase during scattering, 
while 
the states $\ket{\pm}$ and $\ket{\pm i}$
contribute equally.
For p$_{\rm lab}$ above $\approx 100$ MeV (which corresponds to a momentum $p^* \approx 50$ MeV in the center-of-mass frame) mixing to the $\diii$ wave turns on. The stabilizer states $\ket{0}$ and $\ket{1}$ acquire non-vanishing magic during scattering, and the contributions from $\ket{\pm}$ and $\ket{\pm i}$ slightly separate.

As stated in the main text, in the case of NN and YN scattering in the S-wave channels ($\si$ and $\siii$), which was mapped to two qubits, we found that the 60 stabilizers could be organized into groups of states which contribute exactly in the same way to the magic power, or to the entanglement power, of the S-matrix.
The contributions for NN scattering were shown in 
the main manuscript
and those for YN scattering are shown in Figs.~\ref{fig:YN_contrib}. The corresponding groups of stabilizers, which are the same for NN and YN, are detailed below.

\begin{itemize}
\item Group 1 (shown in green in Fig.~\ref{fig:YN_contrib}): this group contains the six tensor-product states where both qubits are in the same state:
\begin{align}
    \ket{0}\otimes \ket{0}  , \; 
    \ket{1}\otimes \ket{1} , \; 
    \ket{+}\otimes \ket{+} , \; 
    \ket{-}\otimes \ket{-} , \; 
    \ket{+i}\otimes \ket{+i} , \; 
    \ket{-i}\otimes \ket{-i} , \; 
\end{align}
which are states number 33, 36, 1 , 4, 17 and 20 in Table~\ref{tab:TwoQstabs},
as well as ten entangled states\footnote{As a reminder, we only include the contributions of both unentangled (tensor-product) and entangled stabilizer states to the magic power, while the entanglement power is only averaged over unentangled stabilizer states.} (number 37, 38, 39, 40, 41, 44, 45, 48, 57 and 60 in Table~\ref{tab:TwoQstabs}), which include the three spin-triplet Bell states, and the singlet. 
These stabilizers lead to outgoing states with no magic, the tensor-product ones also yield no entanglement after scattering. This is true for both NN and YN scattering.

\item Group 2 (shown in blue in Fig.~\ref{fig:YN_contrib}): this group contains the six tensor-product stabilizer states which combines the two eigenstates of the same Pauli operator
\begin{align}
\ket{0}\otimes \ket{1} , \;  
    \ket{1}\otimes \ket{0} , \;  
    \ket{+}\otimes \ket{-} , \; 
    \ket{-}\otimes \ket{+} , \; 
    \ket{+i}\otimes \ket{-i} , \;  
    \ket{-i}\otimes \ket{+i} , \; 
\end{align}
which are states number 34, 35, 2, 3, 18 and 19 in Table~\ref{tab:TwoQstabs},
as well as six entangled states (number 42, 43, 46, 47, 58 and 59 in Table~\ref{tab:TwoQstabs}).
These stabilizer states are particularly interesting in the case of $pn$ and $\Sigma^- n$ scattering, as they scatter into states with large entanglement and vanishing magic at certain energies. For $\Lambda p$ scattering, entanglement and magic follow the same trend.

\item Group 3 (shown in red in Fig.~\ref{fig:YN_contrib}): these are all remaining stabilizer states. These scatter into states presenting the largest magic, but smallest non-zero entanglement in $pn$ and $\Sigma^- n$ scattering. Again for $\Lambda p$ scattering, entanglement and magic display the same behaviour.
\end{itemize}

\begin{figure}[!ht]
    \centering
    \includegraphics[width=.62\columnwidth]{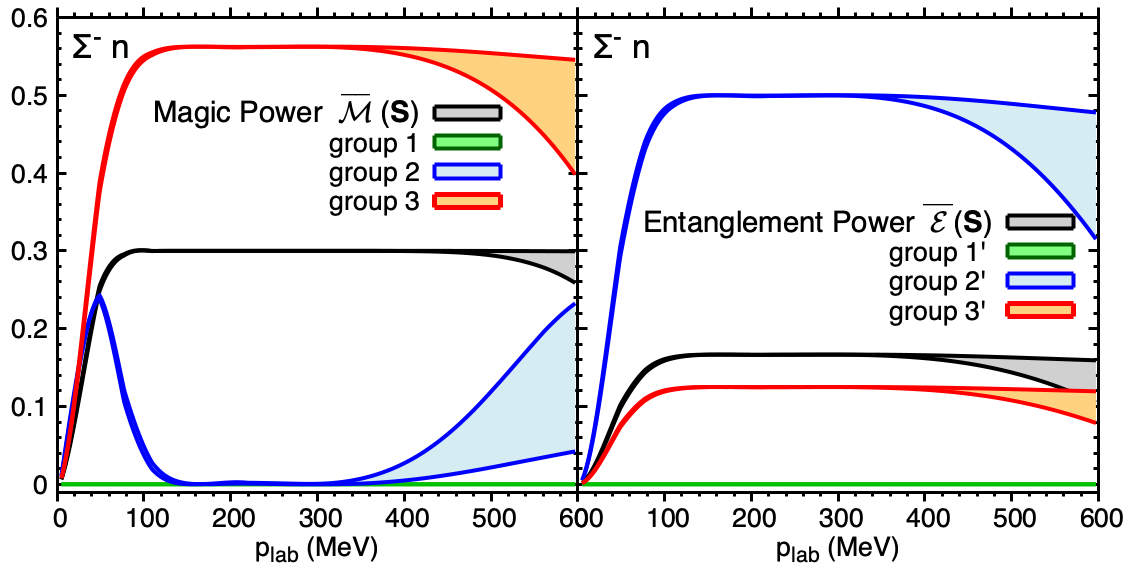}
    \includegraphics[width=.62\columnwidth]{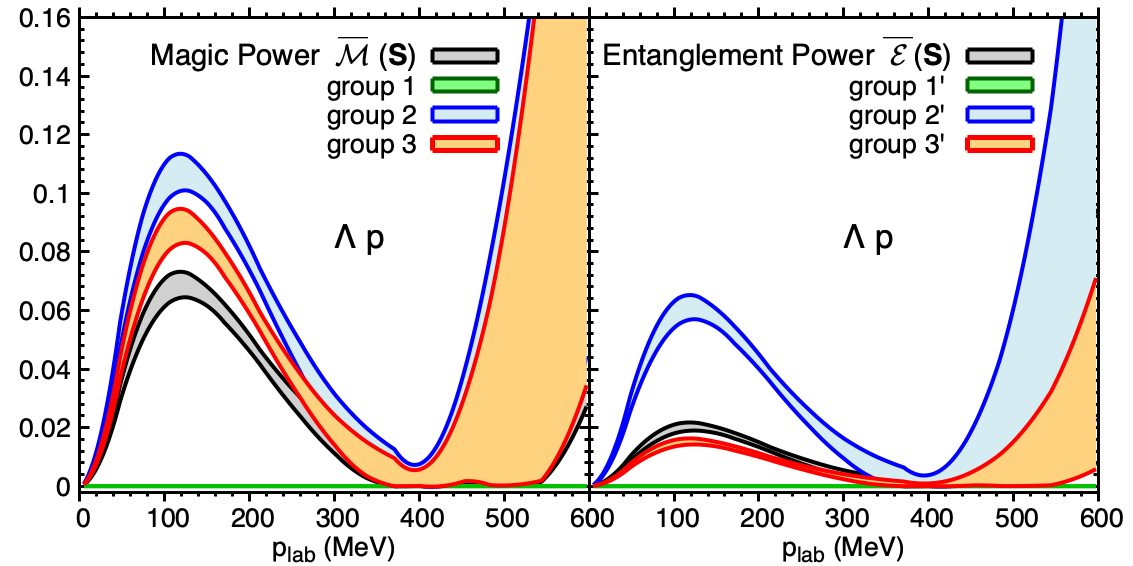}
    \caption{Contributions of different stabilizers states to the magic power $\MS$ and entanglement power $\ES$ of the S-matrix in $\Sigma^- n$ scattering (top two panels) and $\Lambda p$ scattering (bottom two panels). These results have been obtained using N2LO-$\chi$EFT phase shifts from Ref.~\cite{Haidenbauer:2023qhf}. 
    For the magic power, group-1 contains 16 states, group-2 contains 12 states and group-3 contains 32 states.
    For the entangling power,  only the tensor-product stabilizer states of each group are included, (denoted with a prime). Thus, 
    group-1$'$ contains 6 states, group-2$'$ contains 6 states and group-3$'$ contains 24 states.
    We have assumed isospin symmetry between $\Sigma^+p$ and $\Sigma^-n$, and neglected Coulomb interactions.
    The uncertainty bands represent the maximum and minimum values in magic and entanglement derived from the N2LO phase-shift uncertainty bands~\cite{Haidenbauer:2023qhf}.}
    \label{fig:YN_contrib}
\end{figure}
%

\section{Comparison between Chiral-EFT and Phenomenological Phase Shifts}
\label{app:comp_EFT_NSC97}
\begin{figure}[!ht]
    \centering
    \includegraphics[width=.62\columnwidth]{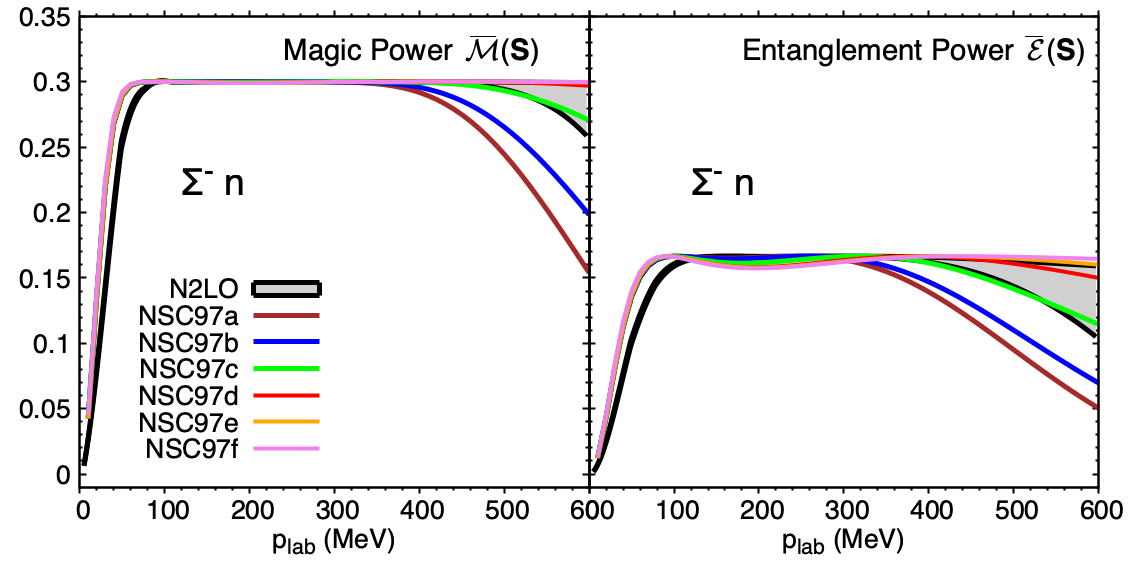}
     \includegraphics[width=.62\columnwidth]{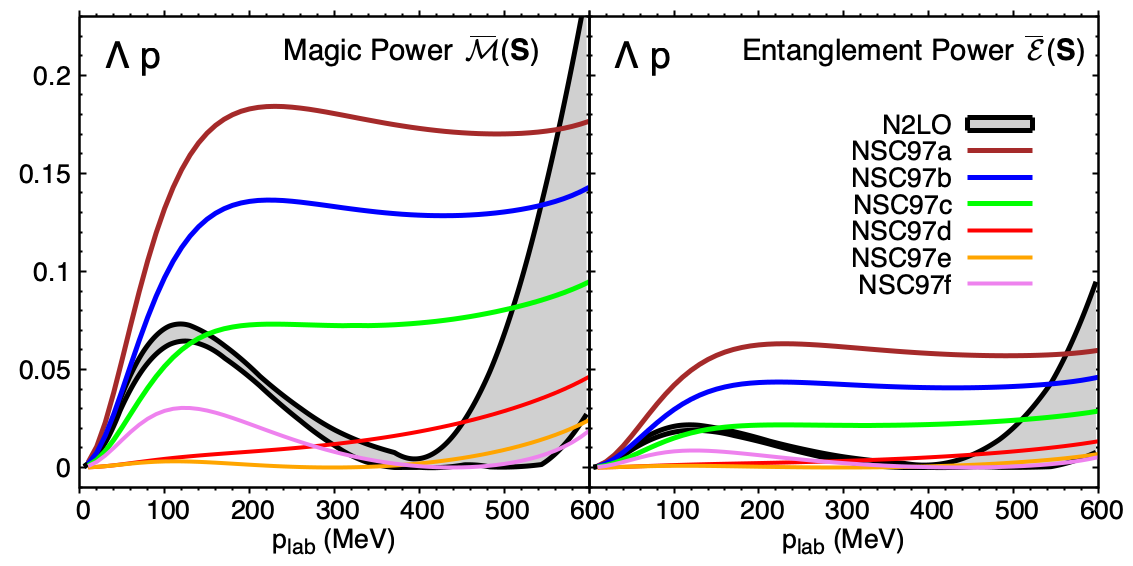}
    \caption{The magic power $\MS$ and entanglement power $\ES$ for $\Sigma^-n$ scattering (top panels) and $\Lambda p$ scattering (bottom panels) obtained from the phase shifts derived from $\chi$EFT of Ref.~\cite{Haidenbauer:2023qhf}, and different parametrizations of the NSC97 phase shifts~\cite{NNonline} (the sign of the NSC97 $^3S_1$ phase shift has been flipped). We have assumed isospin symmetry between $\Sigma^+p$ and $\Sigma^-n$, and neglected Coulomb interactions. The uncertainty bands represent the maximum and minimum values in magic and entanglement derived from the N2LO phase-shift uncertainty bands~\cite{Haidenbauer:2023qhf}.}
    \label{fig:YN_NSC_N2LO_suppl}
\end{figure}
\noindent
While the NN scattering 
phase shifts and mixing parameters are well constrained by experimental data,
the same is only partially true for YN and hyperon-hyperon scattering.
Modern $\chi$EFT analyses of YN scattering, for instance the N2LO results
from Haidenbauer, Mei{\ss}ner, Nogga and Le~\cite{Haidenbauer:2023qhf}, 
provide estimates of the uncertainties determined by the 
expansion parameters of the EFT, defined at leading order in Ref.~\cite{Savage:1995kv}.
Phenomenological potential analyses of the same experimental data sets typically 
do not provide such error estimates.
In the case of the Nijmegen analyses, this lack of error estimates is compensated for by 
an array of different potentials and fits, the NSC97a-f.
For a given observable the  range of predictions from NSC97a-f provides an 
estimate of uncertainty in this phenomenological analysis.

Our results for the magic power and entanglement power in $\Sigma^- n$ and $\Lambda$N that are presented in the main text are derived from the N2LO $\chi$EFT analysis of 
Haidenbauer {\it et al}~\cite{Haidenbauer:2023qhf}.
Here we compare these results and uncertainties from the results obtained using the 
NSC97a-f, an analysis that was state-of-the-art, but constrained by 
less comprehensive data sets, in 1997~\cite{NNonline}.
As discussed in Refs.~\cite{Haidenbauer:2019boi,10.3389/fphy.2020.00012,Haidenbauer:2023qhf}, 
the attractive nature of the phenomenological interactions in the $\Sigma$N ($I=3/2$) $\siii$ channel is inconsistent with recent Lattice QCD calculations~\cite{Beane:2006gf,Beane:2012ey} and empirical information from $\Sigma^-$-formation reactions on nuclei~\cite{Friedman:2007zza}, which point to a repulsive interaction.  
It has been noted that either sign was found to be consistent in the analyses 
that 
led
to NSC97a-f, and a choice was made
in 1997, in the absence of a definitive result, of 
an attractive interaction, but repulsive 
would have also been compatible.
Recent work on entanglement~\cite{liu2023hints}  
uses these $\Sigma^-n$ phase shifts.

Figure~\ref{fig:YN_NSC_N2LO_suppl} shows the magic power and entanglement power in the 
$\Sigma^- n$ and $\Lambda$N channels. According to the above discussion, and for a meaningful comparison, the results in this figure have been obtained by changing the sign of the NSC97a-f phase shifts.
While the results in the $\Sigma^- n$ are consistent between the N2LO analysis and NSC97a-f up to 
p$_{\rm lab}\approx 400$MeV, 
at which point NSC97a and NSC97b become inconsistent with the N2LO error band,
the NSC97a-f predictions in the $\Lambda$N channel 
show little resemblance to the N2L0 prediction and are well outside of the error band 
over most of the energy range.   
In this channel, it is again the case that 
NSC97a and NSC97b are the least consistent.
The magic power and entanglement power in this channel are both small, resulting from substantial cancellations between phase shifts, and this appears to be challenging for the 
NSC97a-f to capture with any precision, which we attribute to the limited data sets available in 1997.


\section{$\Sigma^-$-Catalyzed Magic - Without Decoherence}
\label{app:Sigma_cat}

\noindent In this appendix, we present 
a simple 3-qubit example of the $\Sigma^-$ generating (catalyzing) 
magic between two neutrons by successive scatterings.
Mapping the
$\Sigma^- n n$ spin states to qubits as 
$|s_{\Sigma^-}\rangle \otimes |s_n\rangle\otimes |s_n\rangle$,
the S-matrix for successive scatterings of the $\Sigma^-$ with each neutron is
\begin{eqnarray}
    \hat {\bf S}_{ij} & = &
    \frac{1}{4} \left( 3 \ e^{2 i \delta_1} +  e^{2 i \delta_0} \right ) 
   \hat \Id
    + \frac{1}{4} \left( \ e^{2 i \delta_1} -  e^{2 i \delta_0} \right ) 
    \hat{\boldsymbol{\sigma}_i} .  \hat{\boldsymbol{\sigma}_j}  
    \ \ ,\ \ 
    \hat {\bf S}_{\Sigma^- nn} \ =\  \hat {\bf S}_{13}.\hat {\bf S}_{12}
\label{eq:Smatrix_3q}    
\end{eqnarray}
For a given stabilizer state $|\psi_l\rangle$, 
of which there are 1080, 
we form the scattered state,
$\hat {\bf S}_{\Sigma^- nn}  |\psi_l\rangle$, and then the associated density matrix $\hat \rho^{(l)}$.
To quantify the impact of the $\Sigma^-$ on the two neutrons, the reduced matrix is formed by tracing over the $\Sigma^-$ qubit, leaving the neutron-neutron reduced density matrix $\rho_{nn}^{(l)}$.
This is most easily accomplished by tracing against 3-qubit Pauli strings, and keeping only those of the form $\hat I\otimes\sigma_i\otimes\sigma_j$, with coefficient $c_3^{l;ij}$, and forming $\rho_{nn}^{(l)}$,
\begin{eqnarray}
c_3^{l;ij}  & = & {1\over 8} {\rm Tr}\left[ \hat \rho^{(l)}\  .\  
\hat \Id\otimes\hat \sigma_i\otimes\hat \sigma_j \right]
\ ,\     \rho^{(l)}_{nn} \ =\  2 \sum_{i,j} c_3^{l;ij} \hat \sigma_i\otimes\hat \sigma_j
    \ .
\end{eqnarray}
With the reduced density matrix 
we can apply the same procedure 
to compute the magic as in the main text, but noting that in general $\rho_{nn}^{(l)}$ corresponds to a mixed state and hence the normalizations that were explicit previously 
cannot
be assumed, 
and the probabilities are normalized ``by hand''~\cite{frau2024nonstabilizerness}.
For each stablizer state, the magic $M^{(l)}$ is computed by 
\begin{eqnarray}
c_2^{l;i}  & = & {\rm Tr}\left[ \rho^{(l)}_{nn}. \hat P \right]
\ ,\ 
\Xi^{(l)}_i \ =\ {1\over 4} \left(c_2^{l;i}\right)^2 \nonumber\\
A^{(l)} & = & \sum_i \Xi_i
\ \ ,\ \ 
B^{(l)}\ =\ \sum_i \Xi_i^2 \ \ ,\ \ 
M^{(l)} \ =\   1 - 4 B^{(l)}/A^{(l)}
    \ ,
\end{eqnarray}
and the magic power of the S-matrix acting on the reduced $nn$ system 
(due to successive interactions with the $\Sigma^-$) 
is the average of this ensemble,
$\langle M \rangle \ =\  {1\over N_{\rm stab}}\sum\limits_l M^{(l)} $.

Carrying out this calculation: first setting all phase shifts to zero, gives 
$\langle M \rangle = 0$ as required for stabilizer states, 
and second setting $\Delta\delta=\pi/4$ gives
$\langle M \rangle = 0.405$.
Using $\Delta\delta=\pi/4$ provides a good estimate because the physical phase shift is close to this value and constant over a large energy interval.

It is interesting to note that, of the 1080 3-qubit stablizer states, 
there are only small number of distinct pairs of $\{ A^{(l)}, B^{(l)} \}$ contributing to the average magic power,
\begin{eqnarray}
\{ A^{(l)}, B^{(l)} \} & = & 
\{ 
\{ 1, {1\over 4} \} , 
\{ {7\over 8}, {19\over 128} \} , 
\{ {1\over 2}, {1\over 8} \} , 
\{ {25\over 32}, {259\over 2048} \} , 
\{ {29\over 32}, {227\over 2048} \} , 
\{ {5\over 8}, {11\over 128} \} , 
\{ {17\over 32}, {179\over 2048} \} , 
\{ {17\over 32}, {155\over 2048} \} , 
\nonumber\\
&& \qquad
\{ {21\over 32}, {195\over 2048} \} , 
\{ {21\over 32}, {171\over 2048} \} , 
\{ {25\over 32}, {211\over 2048} \}  
\}
    \ .
\end{eqnarray}

The results of this idealized model of successive $\Sigma^- n$ scatterings 
show that the $\Sigma^-$ induces magic between neutrons.
Given the environment in which these processes are conceivable, 
quantum decoherence due to interactions with other species of particles is expected.
The time-scale of decoherence determines, in part, the impact that the $\Sigma^-$'s will have in 
spreading magic and entanglement at a practical level.

\end{document}